# Mechanical Wave Propagation within Nanogold Granular Crystals

Bowen Zheng[1, 2], Jun Xu[1, 2*]

[1]*Department of Automotive Engineering, School of Transportation Science and Engineering, Beihang University, Beijing, China, 100191*

[2]*Advanced Vehicle Research Center (AVRC), Beihang University, Beijing, China, 100191*

ABSTRACT: We computationally investigate the wave propagation characteristics of nanoscopic granular crystals composed of one-dimensionally arrayed gold nanoparticles using molecular dynamics simulation. We examine two basic configurations, i.e. homogeneous lattices and diatomic lattices with mass-mismatch. We discover that homogeneous lattices of gold nanospheres support weakly dissipative and highly localized solitary wave at 300 K, while diatomic lattices have a good tuning ability of transmittance and wave speed. We establish a validated nonlinear spring contact model with the consideration of complex interactions between gold nanospheres which reveals the physical nature of wave behaviors at nanoscale. This work sheds light on the application of nanogold as a novel mechanical wave tuner, qualitatively and fundamentally different from its counterpart granular materials at meso- and macroscale.

[*]Corresponding author. Tel: +86-10-82339921. Email: junxu@buaa.edu.cn.

# 1. Introduction

Artificially structured phononic crystals and acoustic metamaterials have been at the frontier of science and engineering in recent years due to their novel ability to manipulate mechanical wave propagation in unprecedented ways[1, 2]. The linear response of the periodic system gives rise to many common but useful properties such as the existence of band gaps. However, in the regime of nonlinearity, the wave dynamics in periodic structures become more complex, with no analogs in linear theory. Granular crystals, consisting of tightly packed elastic granules, are a typical example of nonlinear periodic phononic structures, whose nonlinearity rises from the geometry[3, 4]. The simplest form of granular crystals, one-dimensional (1D) chains of identical elastic spheres, were analytically, numerically and experimentally shown to possess the concept of "sonic vacuum" where classical phonons are not supported to highly nonlinear solitary waves (Nesterenko soliton or compacton)[5-8]. The nonlinear response of 1D granular crystals can be tuned in a wide range from linearity, weakly nonlinearity to strongly nonlinearity by the application of a variable precompression[7-9]. Furthermore, by altering the geometry[10-13], material properties[14, 15] and spatial distribution[16-18] of component granules, one is rendered more freedom to manipulate the propagation of mechanical signals, which makes granular crystals building blocks for a broad range of novel applications such as impact-protection devices[13, 19, 20], acoustic diodes[21] and switches[22], and tunable vibration filters[23, 24], just to name a few. Granular crystals composed of granules with diameters at centimeter scale respond to input mechanical signals ranging from 1 Hz to 20000 Hz[3]. However, the frequencies of acoustic waves used in many useful applications such as ultrasonic medical imaging and surgery should reach the order of a megahertz, where downsizing the granules would be a straightforward solution.

Recently, wave propagation in microscopic granular crystals has been experimentally studied[25] and numerical works on nanoscale counterparts (lattices of buckyballs[26-28] and short carbon nanotubes[29, 30]) were conducted via molecular dynamics simulation. It is shown that nanoscopic buckyball granular crystals exhibit an ultra-strong nonlinearity, mainly due to the highly nonlinear molecular repulsive forces instead of the geometry, permitting non-dispersive solitary waves with shorter wavelength and more localized energy compared to its macroscopic counterpart. Exciting as it may seem, the experimental realization of buckyball or carbon nanotube lattices is challenging due to the high-demanding measurement and manipulating equipment, the issue of chemical stability and the requirement of extremely low ambient temperature. Another fascination of granular crystals is the construction of diatomic lattices composed of alternating granules of two different properties such as shape, stiffness and mass etc, adding to the tunability of mechanical waves to a great extent. For example, in compressed diatomic chains of alternating steel and aluminum spherical granules, the existence of discrete breathers was theoretically and experimentally proved, where sustained, exponentially localized oscillations emerge at a frequency inside the forbidden band of the linear Fourier spectrum[31]. For uncompressed diatomic lattices assembled by two different masses, strongly nonlinear resonances and anti-resonances are described by theory and experiments, resulting in an optimal wave attenuation and a new type of solitary wave respectively[32-34].

With the boost of nanotechnology in recent decades, gold nanoparticles have been playing an active role in fields of biology[35], medicine[36], nanocatalysis[37] and metamaterials[38-40] etc. The shape of gold nanoparticles can be well controlled during synthesis[41]. As one of the least reactive chemical elements, gold is an elite candidate for long-term use, especially in corrosive environments. In recent years, there have been plenty of work on the nanogold photonic crystals and electromagnetic metamaterials, exhibiting interesting phenomena with rich physics such as Fano resonance[40] and tunable refractive index[39].

To the best of authors' knowledge, no phononic crystals or acoustic metamaterials at the length-scale of ~$10^2$ nm has been reported. Herein, we report a computational molecular-dynamics (MD) study of nanoscopic nonlinear granular crystals comprising one-dimensionally arrayed gold nanospheres (GNS), one of the thermodynamically stable shapes[42]. We construct two basic configurations of GNS granular crystals, namely, 1D homogeneous GNS lattices, and 1D diatomic GNS lattices with mass-mismatch, simulated at the temperature of 300 K. It is shown that 1D homogeneous GNS lattices support weakly dissipative and highly localized solitary waves at 300 K and the wave properties are quantitatively studied by establishing a nonlinear spring (NS) model. To describe the dynamics of 1D diatomic GNS lattices, the NS model is modified to account for the interaction between two GNSs with different masses. The theoretical predictions of transmission rate and average wave speed are in good agreement with the results of MD simulation.

## 2. Results and discussion

The configurations of 1D homogeneous GNS lattice and diatomic GNS lattice are illustrated in Figure 1, both of which consist of 25 GNSs. The spacing between adjacent GNSs equals to the sum of van der Waals radii, corresponding to the unprecompressed case of their macroscopic counterparts. Diatomic GNS lattices are characterized by a non-dimensional parameter mass ratio $\beta$, defined as the mass ratio of the light and heavy GNS ($0 < \beta \leq 1$). When $\beta = 1$, the diatomic lattice degenerates to a homogenous one; when $\beta$ approaches 0, the lattice becomes a so-called "auxiliary system"[32]. The radii of GNS in homogeneous lattice and heavy GNS in diatomic lattice are both $R_0 = 2.5$ nm. To maintain the contact homogeneity, the 1D lattice is aligned along [001] direction of GNS for the realization of pure facet-facet contact between all neighboring GNSs. The first GNS has a moderate initial velocity $v_{imp}$ ranging from 200 m/s to 1400 m/s, serving as a nanoscale stress wave generator (see Fig. 1). The initial temperature of the system is $T_0 = 300$ K, which strictly prohibits stable solitary wave propagation in 1D buckyball lattices[26].

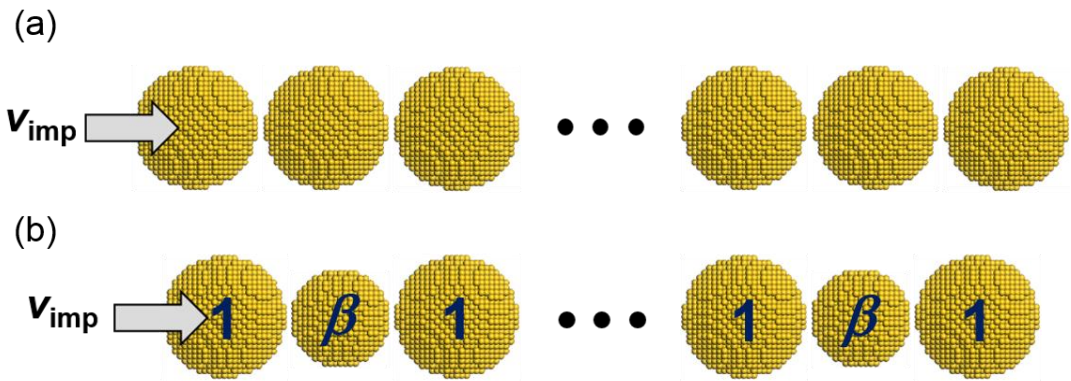

**Fig. 1.** Schematics of 1D nanogold granular crystals. (a) A homogenous GNS lattice. (b) A diatomic GNS lattice.

**2.1. Homogeneous GNS granular crystals.**

For homogeneous GNS lattices, a typical result of wave propagation at $T_0 = 300$ K is shown in Fig. 2, and is compared to low-temperature system at 10 K. To capture the wave behaviors, the resultant force ($F_R$) histories of 6$^{th}$, 9$^{th}$, 12$^{th}$, 15$^{th}$ and 18$^{th}$ GNSs are extracted, normalized as $F_{R,N} = F_R / (E_{Au} R_0^2)$, where $E_{Au}$ = 79 GPa is the Young's modulus of gold. Time is normalized by the time duration of the sound traveling the distance of the diameter, i.e. $t_N = t / (2R_0 / c_{Au})$, where $c_{Au}$ = 2030 m/s is the speed of sound in a gold rod. The moment of impact is shifted to $T_N = 0$. The normalized impact velocity is given as $v_{imp,N} = v_{imp} / c_{Au}$, and three impact velocities applied here are $v_{imp,N}$ = 0.197, 0.296 and 0.394. According to the results shown in Figs. 2(a) and 2(b), stable traveling nonlinear waves are supported by homogeneous GNS lattice at $T_0 = 300$ K, and the comparison to system at 10 K reveals that the behaviors of the traveling wave are not sensitive to the temperature effect, in contrast to buckyball lattices. It is observed that the wave propagation actually has a little attenuation (< 5% from 6$^{th}$ to 18$^{th}$ GNS), corresponding to the concept of weakly dissipative solitary wave in literature[43]. It is obvious from Figs. 2(a) and 2(b) that the waves have a nonlinear property of amplitude-dependent wave speed: the larger the amplitude, the larger the wave speed. Fig. 2(c) shows the change of system temperature upon various impact velocities. The temperature is normalized as $T_N = T / T_0$. Intuitively, higher impact velocity gives rise to higher temperature rise, but the peak value of system temperature is well below the melting point of gold (1337.33 K[44]).

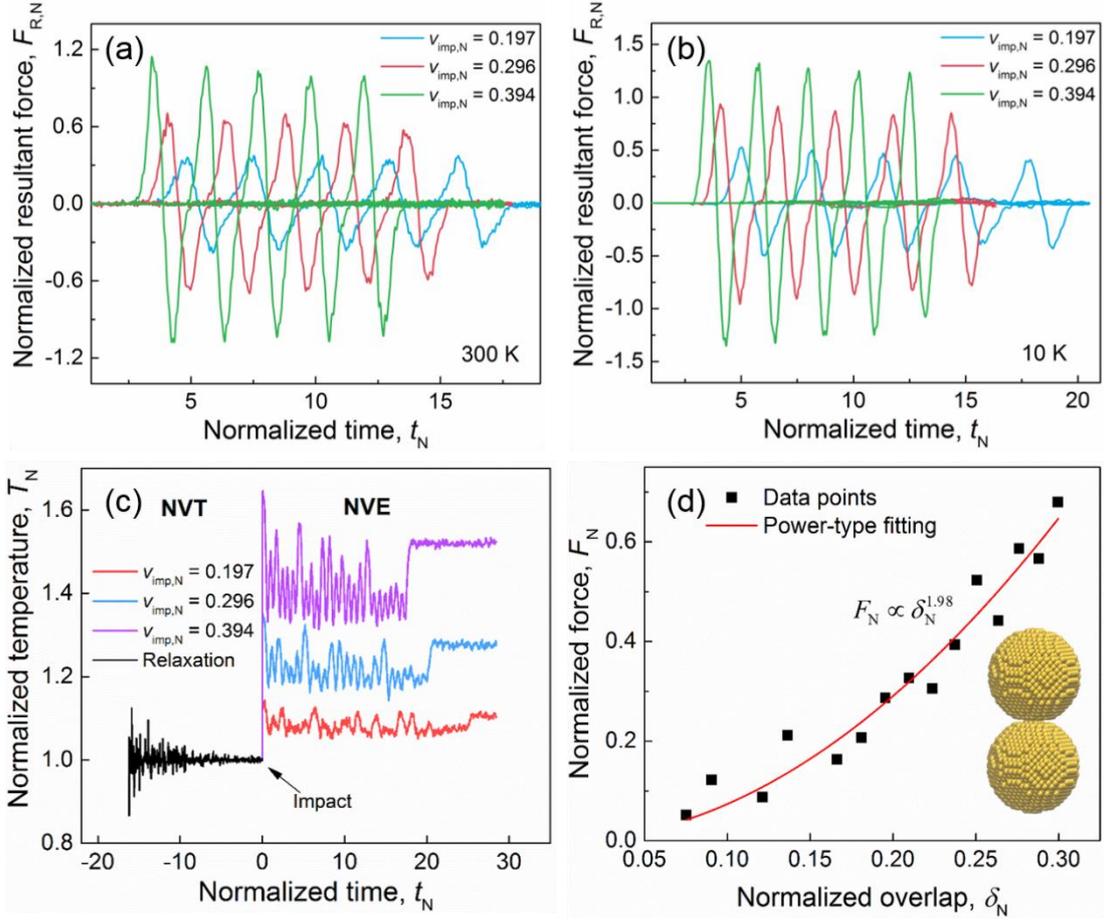

**Fig. 2.** Simulation results of wave propagation in 1D homogeneous GNS lattices. (a-b) Wave profiles at 300 K and 10 K. (c) The change of system temperature due to various impacts. (d) Force-overlap relation from a MD compression test.

At macroscale or microscale, granular crystals can be described as a set of coupled nonlinear oscillators governed by Herzian contact, under the assumptions of small, elastic deformation and frictionless surfaces[45]. The Hertz law gives the compression force-overlapping distance relation of two spheres as

$$F_{ij} = \frac{4}{3}\left(\frac{1-v_i^2}{E_i} + \frac{1-v_j^2}{E_j}\right)^{-1}\left(\frac{R_i R_j}{R_i + R_j}\right)^{1/2} \delta_{ij}^{3/2} \qquad (1)$$

where $\delta_{ij} = (R_i + R_j) - (x_j - x_i)$ is the overlapping distance; $F$ is the compression force; $R$ is the radius; $x$ is the coordinate of the center of the sphere; $E$ is Young's

Modulus; $\nu$ is Poisson's ratio. The subscripts $i$ and $j$ denote two contacting granules. The expression of Eq. (1) is of decisive importance to the wave dynamics in granular crystals. However, for nanoscale contact, the breakdown of continuum theory is demonstrated[46], and the "rough" surfaces of nanoparticles consisting of crystal steps and terraces lead to more complex interaction law. Although many original investigations on nanoscale contact problem has been done[47-49], hitherto no complete analytical result resembling Eq. (1) is available.

To model the contact interaction between nanoparticles for the study of wave propagation, we establish an empirical nonlinear spring (NS) model $F = K\delta^n$, as a reasonable simplification of the complicated interaction between adjacent GNSs, where $F$ is the interaction force; $\delta$ is the overlap distance of van der Waals spheres of two GNSs; $K$ is the stiffness parameter; $n$ is the nonlinear index. NS model has been proved accurate and efficient in modeling the interaction between buckyballs in the context of 1D and 2D stress wave propagation[26, 28]. By impacting two GNSs in [001] direction at $T_0$ = 300 K and fitting the force-overlap data points in a power-type function $F_N = K^* \delta_N^n$, where $F_N = F/(E_{Au} R_0^2)$ and $\delta_N = \delta/(2R_0)$, we obtain the NS parameters $K = 3.73 \times 10^{11}$ and $n = 1.98$ (see Fig. 2(d)). Note that for macroscale sphere-sphere Herzian contact, the nonlinear index is 1.5. This suggests that the repulsive force between two GNSs indeed has a stronger nonlinearity compared to its macroscale counterpart, further indicating that the existence of more spatially localized nonlinear waves is possible.

The governing equations of motion of 1D homogeneous GNS lattices based on NS

model can be given as

$$M\frac{d^2u_1}{dt^2} = -K[\delta_2]_+^n - \mu\frac{d\delta_2}{dt}\Theta(\delta_2);$$

$$M\frac{d^2u_i}{dt^2} = K\left([\delta_i]_+^n - [\delta_{i+1}]_+^n\right) + \mu\left(\frac{d\delta_i}{dt}\Theta(\delta_i) - \frac{d\delta_{i+1}}{dt}\Theta(\delta_{i+1})\right), 1<i<N; \quad (2)$$

$$M\frac{d^2u_N}{dt^2} = K[\delta_N]_+^n + \mu\frac{d\delta_N}{dt}\Theta(\delta_N).$$

$$\delta_i = u_{i-1} - u_i, N = 25$$

where $M$ is the mass of a GNS; $u_i$ is the displacement of $i^{th}$ GNS; The subscript (+) is defined by $[r]_+ = \max(0,r)$; $\mu$ is a dimensional damping coefficient, introduced to describe the weak attenuation; $\Theta(\bullet)$ is the Heaviside function, used to account for the fact that damping force are only active during compression and are absent when two GNSs separate. A normalization is introduced as

$$x_i = \left(\frac{4\pi}{3}\right)^{\frac{1}{n-1}}\frac{1}{R}u_i;$$

$$\tau = \frac{3}{4\pi}\left(\frac{K}{\rho}R^{n-4}\right)^{\frac{1}{2}}t = \psi t; \quad (3)$$

$$\lambda = \mu/(M\psi) = \left(\rho K R^{n+2}\right)^{-\frac{1}{2}}\mu.$$

where $x_i$, $\tau$ and $\lambda$ are non-dimensional displacement, time and damping coefficient respectively and $\rho$ is the mass density of gold. Thus, a set of non-dimensional governing equations of motion can be obtained as

$$\frac{d^2x_1}{d\tau^2} = -[\Delta_2]_+^n - \lambda\frac{d\Delta_2}{d\tau}\Theta(\Delta_2);$$

$$\frac{d^2x_i}{d\tau^2} = [\Delta_i]_+^n - [\Delta_{i+1}]_+^n + \lambda\left(\frac{d\Delta_i}{d\tau}\Theta(\Delta_i) - \frac{d\Delta_{i+1}}{d\tau}\Theta(\Delta_{i+1})\right), 1<i<N; \quad (4)$$

$$\frac{d^2x_N}{d\tau^2} = [\Delta_N]_+^n + \lambda\frac{d\Delta_N}{d\tau}\Theta(\Delta_N).$$

$$\Delta_i = x_{i-1} - x_i, N = 25$$

Applying a weak damping term by setting $\lambda = 0.007$, the detailed wave behaviors of 1D homogeneous GNS chain upon different impact velocities ($v_{\text{imp,N}}$ = 0.197 and 0.296) can be precisely captured by the NS model, as is shown in Fig. 3.

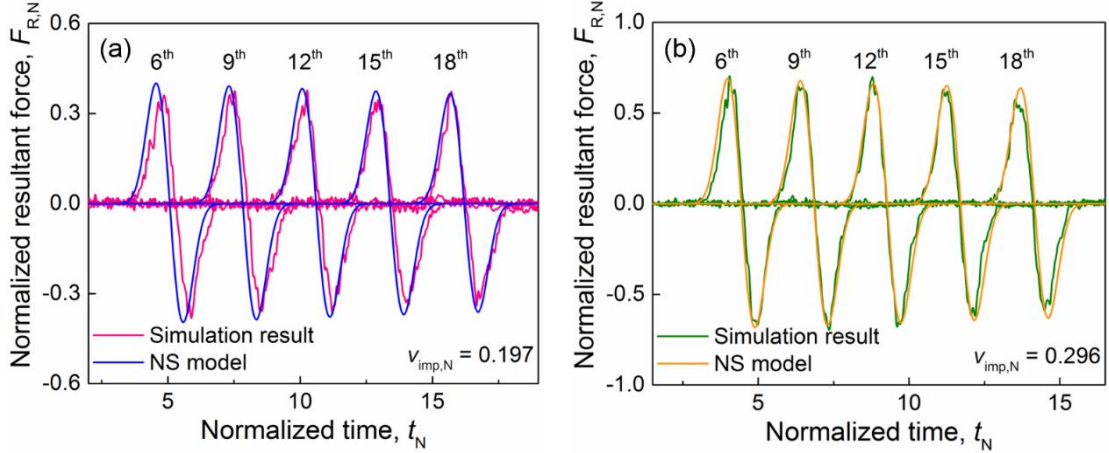

**Fig. 3.** Comparison between the results of MD simulation and numerical calculation based on Eq. (3) upon different impact velocities. (a) $v_{\text{imp,N}}$ = 0.197. (b) $v_{\text{imp,N}}$ = 0.296.

The properties of the waves in 1D homogeneous GNS granular crystals are discussed as follows. Nesterenko showed that an unprecompressed granular crystal characterized by a generalized power-type nonlinear contact permits the formation and propagation of strongly nonlinear solitary wave[8], the relation between wave speed and amplitude is given as

$$c \propto A^{\frac{n-1}{2n}} \qquad (5)$$

where $c$ is wave speed and $A$ is force amplitude. The "scaling relation" between wave speed and maximum particle velocity is

$$c \propto v_{\text{Pmax}}^{\frac{n-1}{n+1}} \qquad (6)$$

where $v_{\text{Pmax}}$ is the maximum particle velocity. For an ideal highly nonlinear solitary wave, wave speed, amplitude and maximum particle velocity are constant as the wave propagates, while in homogeneous GNS lattices these quantities can be slightly lowered after the wave has traveled a distance due to the existence of weak damping. Therefore, the force amplitude and maximum particle velocity are taken as the average of 6th, 9th, 12th, 15th and 18th GNSs, and the wave speed is calculated dividing the distance between 6th and 18th GNSs by the time difference for the two GNSs to reach resultant force maxima. For the nonlinear index $n$ = 1.98, Eqs. (4) and (5) give $c \propto A^{0.25}$ and $c \propto v_{Pmax}^{0.33}$. These relations are validated in Figure 4, where normalizations are conducted as $c_N = c/c_{Au}$, $v_{Pmax,N} = v_{Pmax}/c_{Au}$ and $A_N = A/(E_{Au} R_0^2)$.

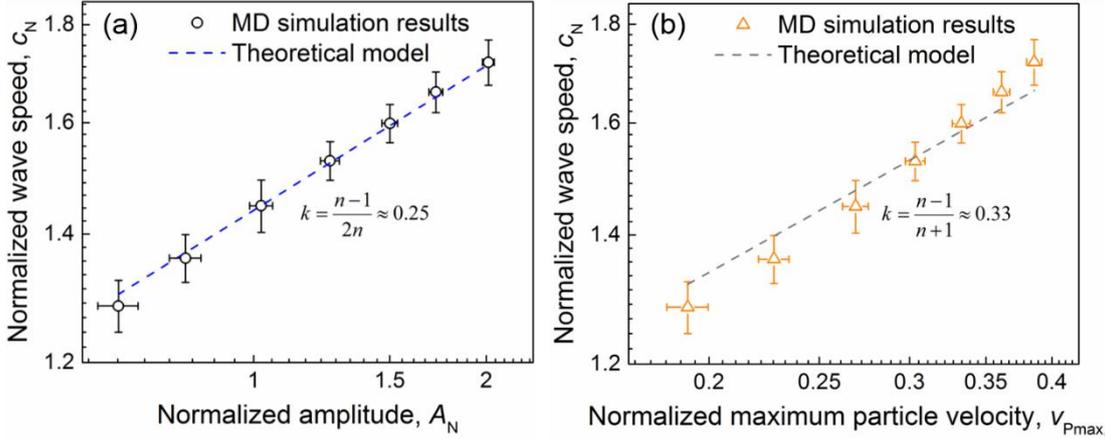

**Fig. 4.** Validation of the nonlinear wave properties presented by Eqs. (4) and (5). Theoretical model is characterized by the slope $k$ in semi-log graphs. (a) The relation between wave speed and amplitude, where $k \approx 0.25$. (b) The relation between wave speed and maximum particle velocity, where $k \approx 0.33$.

Theoretically, waves in granular crystals of power-type contact with no precompression are dispersionless with a constant wave width, whose dependence on $n$ ($n > 1$) is expressed as[8]

$$L = \frac{\pi a}{n-1}\sqrt{\frac{n(n+1)}{6}} \qquad (7)$$

where $a$ is the distance between the geometric centers of adjacent particles, equal to the equilibrium spacing between adjacent GNSs. A nonlinear index $n$ = 1.98 outputs $L \approx 3a$, consistent with the simulation result in Figure 5 where waves are localized within a width of 4 consecutive GNSs as propagation. In Figure 5, the particle velocities of all GNSs are extracted at three instants, i.e. $t_N$ = 4.06, 8.93 and 13.80. The impact velocity is set as $v_{imp,N}$ = 0.296 and particle velocity is normalized as $v_{P,N} = v_P / c_{Au}$. Above observations confirm that unprecompressed 1D homogeneous GNS granular crystals are able to support dissipative Nesterenko solitary wave at room temperature, making the experimental realization of nanoscale solitary wave promising.

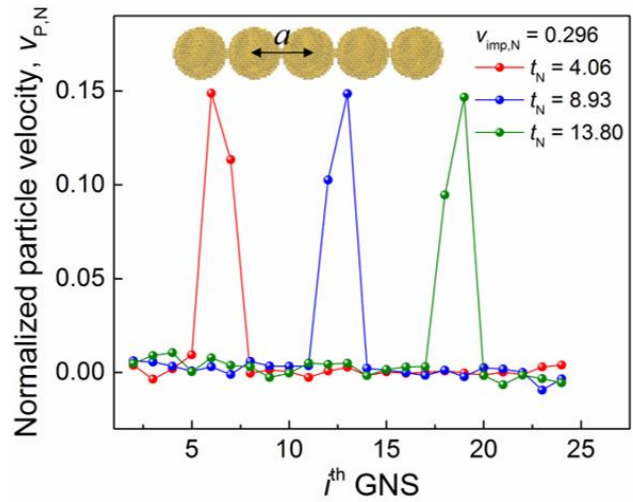

**Fig. 5.** Particle velocities of all GNSs at three instants, i.e. $t_N$ = 4.06, 8.93 and 13.80.

## 2.2. Diatomic GNS granular crystals.

1D uncompressed diatomic granular crystals composed of alternating heavy and light GNSs are constructed and studied. The set of non-dimensional governing equations of motion can be given as

$$\frac{d^2 x_1}{d\tau^2} = -[\Delta_2]_+^n - \lambda \frac{d\Delta_2}{d\tau}\Theta(\Delta_2);$$

$$\beta \frac{d^2 x_i}{d\tau^2} = [\Delta_i]_+^n - [\Delta_{i+1}]_+^n + \lambda\left(\frac{d\Delta_i}{d\tau}\Theta(\Delta_i) - \frac{d\Delta_{i+1}}{d\tau}\Theta(\Delta_{i+1})\right);$$

$$\frac{d^2 x_{i+1}}{d\tau^2} = [\Delta_{i+1}]_+^n - [\Delta_{i+2}]_+^n + \lambda\left(\frac{d\Delta_{i+2}}{d\tau}\Theta(\Delta_{i+2}) - \frac{d\Delta_{i+3}}{d\tau}\Theta(\Delta_{i+3})\right); \quad (8)$$

$$i = 2, 4, 6 \ldots$$

$$\frac{d^2 x_N}{d\tau^2} = [\Delta_N]_+^n + \lambda \frac{d\Delta_N}{d\tau}\Theta(\Delta_N).$$

$$\Delta_i = x_{i-1} - x_i, \quad N = 25$$

In diatomic lattices, the contact problem is even more complex. An important indication of Herzian contact expressed in Eq. (1) is that the nonlinearity of sphere-sphere Hertzian contact is independent of the geometry and material properties, i.e. $R$, $E$ and $\nu$, making $\beta$ the only non-dimensional governing parameter in Eq. (8). However, due to the complexity of the contact law between nanospheres, the simple relation of $F \propto \delta^n$ may no longer hold when two nanoparticles with different masses collide, and the contact nonlinearity is naturally considered as a function of geometric parameters. Assuming the mass density of gold is constant, we have $\beta = (R_L / R_H)^3$, where $R_L$ and $R_H$ refer to the radii of light and heavy GNSs respectively. Now, the aim becomes establishing the functional relation between interaction nonlinear index $n$ and mass ratio $\beta$, i.e. $n = n(\beta)$.

In this study, the strategy of achieving various mass ratios is varying the mass of light GNS while keeping the heavy GNS unaltered. Setting $R_H \equiv R_0 = 2.5$ nm, an efficient radius $R_{\text{eff}}$ is defined as

$$R_{\text{eff}}(\beta) = \frac{2 R_H R_L}{R_H + R_L} = \frac{2\beta^{\frac{1}{3}}}{1 + \beta^{\frac{1}{3}}} R_H \quad (9)$$

which homogenizes the different radii of the two interacting GNSs, as is illustrated in Fig. 6(a). Due to the small size of GNS, the mechanism of repulsive force can fundamentally deviate from macroscale Hertzian contact, and may be mainly accounted for by Pauli repulsion at short ranges[50]. Therefore, the potential between GNSs is modeled by an empirical 12-6 potential[51]

$$V(r,\beta) = C\left[\left(\frac{r_m(\beta)}{r}\right)^{12} - 2\left(\frac{r_m(\beta)}{r}\right)^6\right] \quad (10)$$

$$r_m(\beta) = 2R_{\text{eff}}(\beta)$$

where $r$ is the distance between the centers of two interacting GNSs; $r_m(\beta)$ is the equilibrium spacing, as a function of mass ratio $\beta$; $C$ is a constant. The interaction force can be then obtained by taking the partial derivatives of potential $V(r,\beta)$ with respect to $r$, i.e. $F(r,\beta) = -\partial V(r,\beta)/\partial r$. To simplify the problem and pursue a sound physical indication, a modified NS model is put forward

$$F = K\delta^{n(\beta)} \quad (11)$$

It is calibrated by the force-displacement curve derived by the 12-6 model in Eq. (10). Part of the calibration are shown in Fig. 6(b), where the modified NS model in Eq. (11) is fit with the derivatives of Eq. (10) from 0 to $\delta = 0.25R_0$ (corresponding to maximum deformation possible) of various $\beta$ to obtain corresponding $n$. For $\beta = 1, 0.8, 0.6, 0.4$, the power-type fitting gives $n = 2.03, 2.09, 2.16, 2.29$. The accuracy of calibrating $n$ as a function of $\beta$ is validated by impacting heavy and light GNSs of various mass ratios at $T_0 = 300$ K, as is shown in Fig. 6(c). The dashed line is the result of calibrated modified NS model, whose precision is acceptable. It is observed that $n$ is a decreasing function of $\beta$, but with a slowing-down rate, suggesting that for nanoparticles, contact nonlinearity can be increased by creating mass difference.

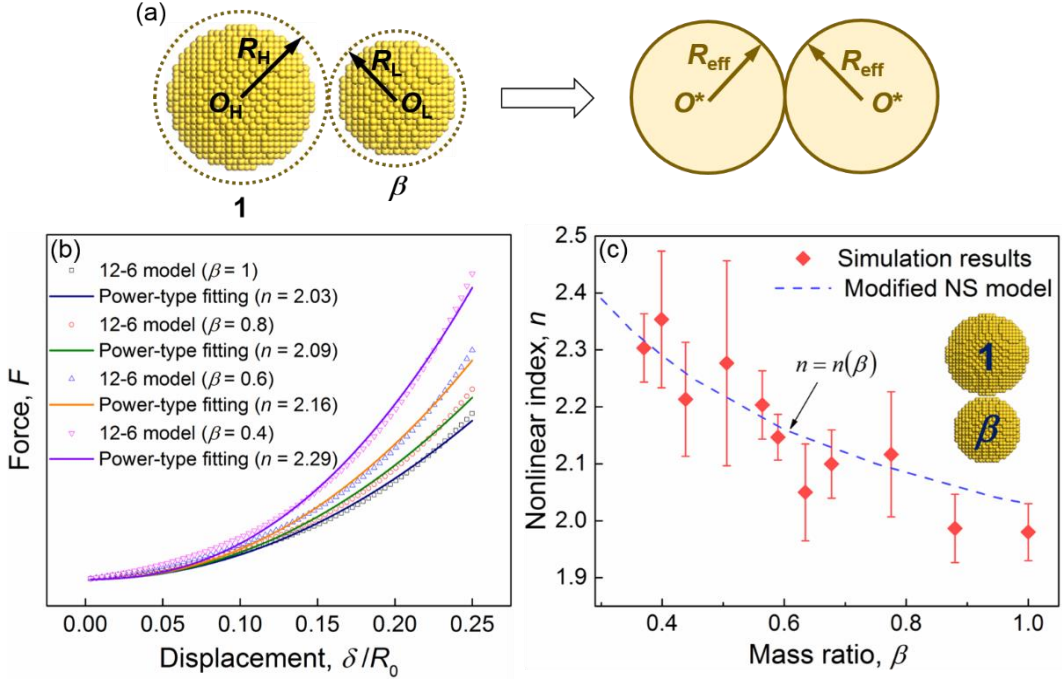

**Fig. 6.** The establishment of interaction law between GNSs. (a) Illustration of effective radius $R_{eff}$, where $R_H$ and $R_L$ are the radii of heavy and light GNSs; $O_H$, $O_L$ and $O^*$ are the centers of heavy, light and effective GNSs respectively. (b) Calibrating modified NS model by the 12-6 model. The dots represent the results of 12-6 model and the solid lines are fitting results. (c) Validation of the calibration by impacting heavy and light GNSs of various mass ratios at $T_0 = 300$ K.

It is shown in literature that nonlinear resonances in uncompressed diatomic granular crystals result in a significant wave attenuation, which, from a practical point of view, could be employed as passive shock attenuators[32]. A transmission rate $\eta$ is defined as $\eta = v_{P\max,last} / v_{imp}$ to evaluate the attenuation performance of the diatomic GNS granular crystals, where $v_{P\max,last}$ is the maximum particle velocity of the last GNS, Different from macroscale diatomic lattices governed by Hertzian contact, governing Eq. (7) indicates that the dynamics of GNS diatomic granular crystals is a complicated interplay of mass ratio, nonlinearity and damping levels, i.e.

$\eta = \eta(\beta, n, \lambda)$. For various damping levels ($\lambda = 0$ (undamped), 0.005, 0.01, 0.015, 0.02), given a set of ($\beta$, $n$), $\eta$ is obtained by numerically solving Eq. (7) using the fourth order Runge-Kutta method, presented in the form of $\beta$-$n$ maps in Fig. 7(a). It is observed that for arbitrary $n$ from 1.5 to 2.5, the resonance zone always occurs at $0.5 < \beta < 0.6$. However, larger $n$ results in lower $\eta$ at resonance zone. This suggests that theoretically, the position of resonance zone does not change with contact nonlinearity, while the attenuation induced by nonlinear resonance is intensified when the nonlinearity gets stronger. In addition, enhanced damping level will globally lower the transmittance, in other words, improve the attenuation ability but will not shift the position of resonance zone. The numerical results in Fig. 7(a) may serve as a theoretical guidance to the design of granular crystals to obtain desired transmittance of mechanical signals as well as optimal attenuation. For the interaction in diatomic GNS lattices, the one-to-one relation between $\beta$ and $n$ has been empirically established in modified NS model. Therefore, the system dynamics should lie on a trajectory on $\beta$-$n$ maps. Transforming the dashed curve in Fig. 6(c) to the $\beta$-$n$ map, the trajectory of the system's dynamics is visualized as the black curve in Fig. 7(a). Thus, for a fixed damping level, $\eta$ could be theoretically predicted for any given $\beta$. The accuracy of this prediction is compared to the MD simulation results upon various impact velocities in Fig. 7(b). It is found that the position of resonance zone can be precisely predicted. Note that only a range of transmission rate could be given because the damping level is also a function of $\beta$. At the resonance mass ratio, the transmission rate can be as low as ~0.3, rendering the diatomic GNS lattice the potential to serve an effective nanoscale impact protector.

**Fig. 7.** Transmission rate as a function of mass ratio $\beta$, nonlinear index $n$ and non-dimensional damping coefficient $\lambda$. (a) $\beta$-$n$ maps for $\lambda = 0, 0.005, 0.01, 0.015, 0.02$. The black trajectory is the functional relation $n = n(\beta)$ of modified NS model. (b) Validation of modified NS model with MD simulation results upon the impact velocities of $v_{imp,N} = 0.197, 0.296$ and $0.394$ at $T_0 = 300$ K.

It has been reported that in 1D diatomic granular lattices at macroscale, for an infinite number of mass ratios, inherent nonlinear resonances or anti-resonances could happen, resulting in a maximum wave attenuation or a new family of solitary respectively. In Figure 8, the resultant force histories of a light GNS in diatomic lattices ($\beta = 0.5$) and a GNS in homogeneous lattices are extracted. For the light GNS in diatomic lattices, both a squeeze mode and a collision mode are observed, while for the GNS in homogeneous lattices, there appears no oscillating tails. These wave characteristics are in good accordance with theoretical descriptions of nonlinear resonances and anti-resonances in Ref. [32-33].

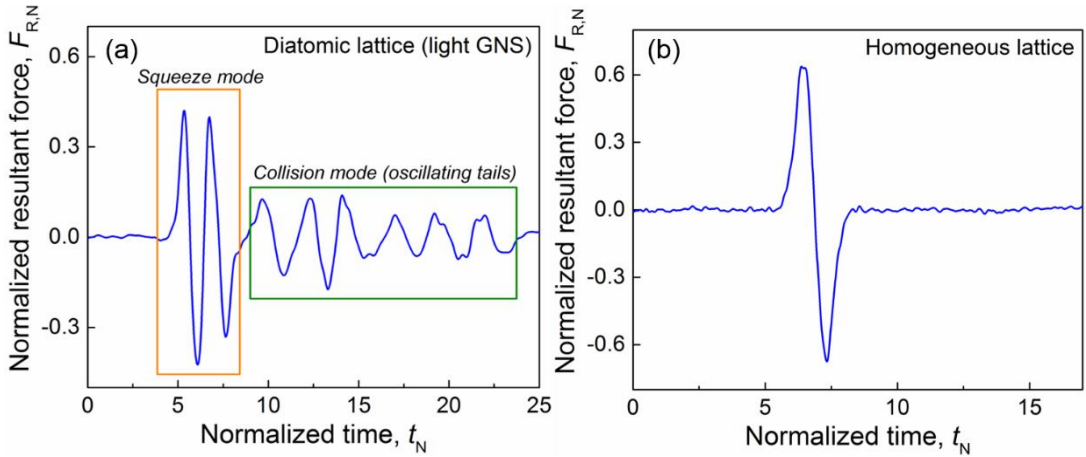

**Fig. 8.** Resultant force histories of a single GNS upon $v_{imp,N} = 0.296$ at $T_0 = 300$ K. (a) A light GNS in diatomic lattices ($\beta = 0.5$). (b) A GNS in homogeneous lattices.

Having examined the wave transmission property of diatomic GNS granular crystals, we next focus on how average wave speed $c_{avg}$ is influenced by these governing factors, i.e. the functional relation of $c_{avg} = c_{avg}(\beta, n, \lambda)$. The average wave speed is calculated based on the length of entire 1D GNS lattices. Similarly, for various damping levels, given a set of ($\beta$, $n$), $c_{avg,relative}$ is obtained by numerically solving Eq. (7), as is shown in the form of $\beta$-$n$ map in Figure 9(a). A relative average wave speed $c_{avg,relative}$ is applied here due to the lack of stiffness parameters for various $\beta$. From the numerical result, it is observed that the position of slowest average wave speed almost coincides with the resonance zone for transmittance, and that $c_{\text{avg,relative}}$ gets lower with increasing $n$. Additionally, enhanced damping level will globally slow down the average wave speed, but will not shift the position of the slowest wave speed with respect to $\beta$. This result may inspire the design of controlling the flowing rate of mechanical energy using granular crystals. According to the one-to-one relation between $\beta$ and $n$ established in the modified NS model, the prediction of relative average wave speed of diatomic GNS lattices lies on the black curves on the $\beta$-$n$ maps in Fig. 9(a). The results of MD simulation upon different impact velocities are presented in Fig. 9(b). As can be seen, larger impact velocities outputs larger average wave speed, as the nonlinear dynamics of GNS granular crystals suggest. The position of the slowest wave speed is unaltered with upon different impact velocities and coincides with theoretical predictions in Fig. 9(a).

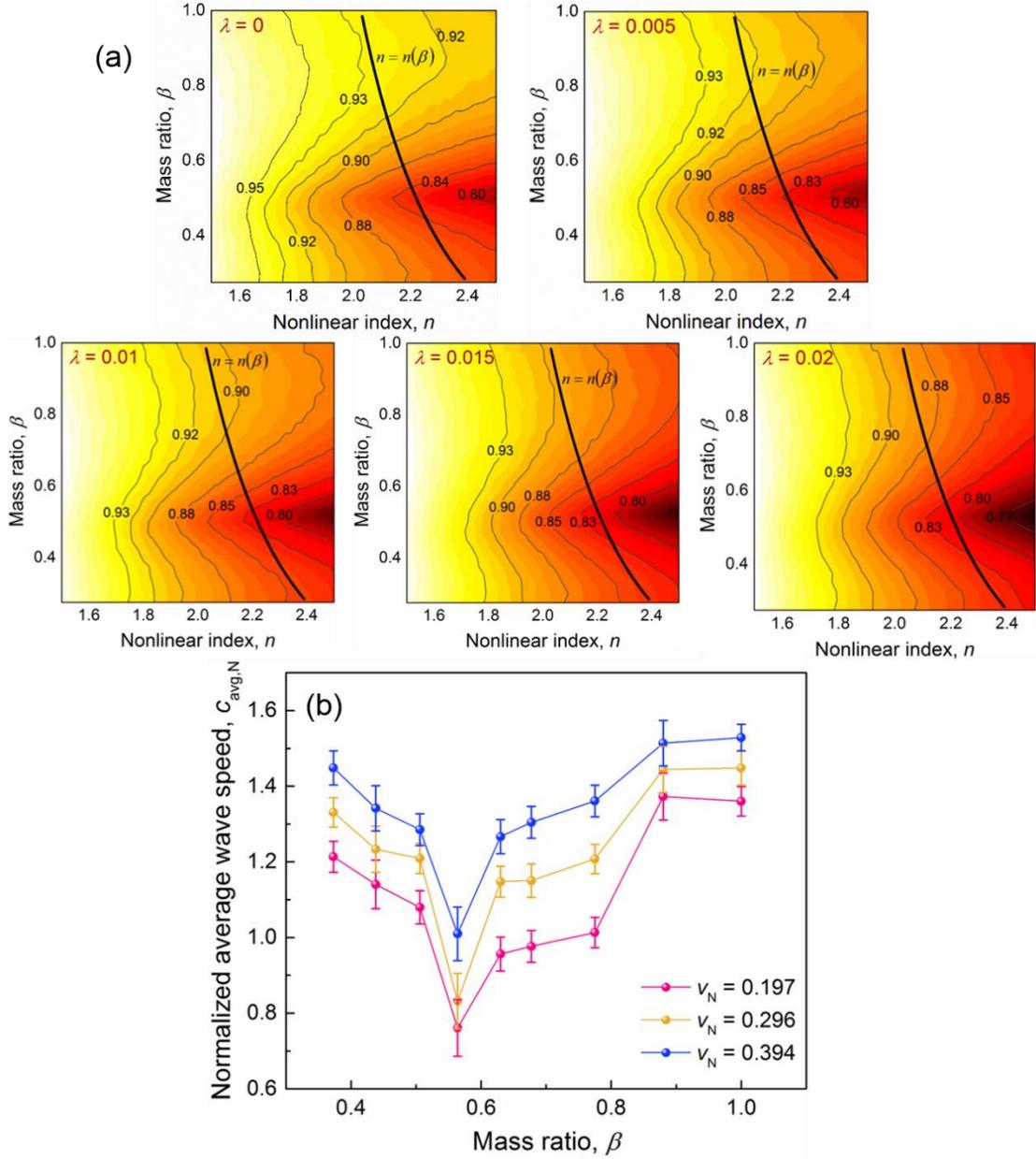

**Fig. 9.** Average wave speed as a function of mass ratio $\beta$, nonlinear index $n$ and non-dimensional damping coefficient $\lambda$. (a) $\beta$-$n$ maps for $\lambda$ = 0, 0.005, 0.01, 0.015, 0.02. The black trajectory is the functional relation $n = n(\beta)$ of modified NS model. (b) MD simulation results upon various impacts, i.e. $v_{imp,N}$ = 0.197, 0.296 and 0.394.

## 3. Conclusions

We have computationally studied the wave propagation characteristics of nanoscopic granular crystals composed of one-dimensionally arrayed gold nanoparticles using MD simulation. We have found that 1D homogeneous GNS lattices support the formation and propagation of weakly dissipative and highly localized solitary wave at room temperature, while 1D diatomic GNS lattices have a good tuning ability of transmittance and wave speed due to the phenomenon of nonlinear resonances. We have established an empirical NS model and made necessary modifications to accurately describe the wave dynamics of both systems. Although the deviation from Herzian contact makes interaction law between nanoparticles more complex, it renders nanoscale granular crystals the potential to obtain a broad, continuously alterable nonlinearity, which will possibly inspire the design of granular crystals beyond traditional functions. To this end, an analytical model of nanoscale contact is called for. Thanks to the great downsizing of the granules, the GNS granular crystals should be able to respond to mechanical waves of ultrahigh frequency, and hopefully can pave the way for the development of applications such as ultrasonic medical imaging and nondestructive detection in solids. We anticipate that this work can shed light on the application of nanogold as a mechanical wave tuner that is qualitatively different from its macroscale counterparts, and offer the exciting possibility of its experimental realization.

## Acknowledgements

This work is financially supported by Fundamental Research Funds for the Central Universities, Beihang University and start-up funds of "The Recruitment Program of Global Experts" awardee (YWF-16-RSC-011), Beihang University.

## Appendix: Molecular dynamics simulation

The molecular dynamics simulations are conducted based on the open-source program LAMMPS (Large-scale Atomic/Molecular Massively Parallel Simulator)[52], and visualized using VMD (Visual Molecular Dynamics)[53]. The potential between gold atoms are computed by a Lennard-Jones (L-J) 6-12 pairwise potential, which is widely used for simulating metal particles:

$$U(r_{ij}) = 4\varepsilon\left[\left(\frac{\sigma}{r_{ij}}\right)^{12} - \left(\frac{\sigma}{r_{ij}}\right)^{6}\right] \quad (12)$$

where $U$ is the L-J potential for two atoms; $r_{ij}$ is the distance between atom $i$ and $j$; $\varepsilon$ and $\sigma$ are two L-J parameters, representing the depth of the potential well and the finite distance for zero the inter-particle potential respectively, chosen as $\varepsilon = 0.458\,\text{eV}$ and $\sigma = 2.569\,\text{Å}$ [54]. A spherical cutoff is taken to be $2.2\sigma$[55, 56]. A periodic boundary condition is applied in all dimensions. The time integration step is set as 1 femtosecond. The system first runs for equilibrium in NVT ensemble (the canonical ensemble) at temperature $T_0 = 300$ K for 40000 fs and then runs for 80000 fs in NVE ensemble (the micro-canonical ensemble) for the investigation of wave propagation. Lateral resultant forces are eliminated to maintain the one-dimensionality during the simulation.